\documentstyle[11pt,aaspp4]{article}

\def\la{\mathrel{\hbox{\rlap{\hbox{\lower4pt\hbox{$\sim$}}}\hbox{$<$}}}}
\def\ga{\mathrel{\hbox{\rlap{\hbox{\lower4pt\hbox{$\sim$}}}\hbox{$>$}}}}
\def\etal{et al.\,\,}

\slugcomment{Accepted to {\it The Astrophysical Journal}}

\begin{document}

\title{A Deficit of High-Redshift, High-Luminosity X-Ray Clusters:\nl
Evidence for a High Value of $\Omega_m$?}

\author{D. E. Reichart\altaffilmark{1,2}, R. C. Nichol\altaffilmark{2}, F. J. Castander\altaffilmark{1}, D. J. Burke\altaffilmark{3}, \nl 
A. K. Romer\altaffilmark{2}, B. P. Holden\altaffilmark{1}, C. A. Collins\altaffilmark{3}, M. P. Ulmer\altaffilmark{4}}

\altaffiltext{1}{Department of Astronomy and Astrophysics, University of Chicago, 5640 South Ellis Avenue, Chicago, IL 60637} 
\altaffiltext{2}{Department of Physics, Carnegie Mellon University, 5000 Forbes Avenue, Pittsburgh, PA 15213}
\altaffiltext{3}{Astrophysics Research Institute, Liverpool John Moores University, Liverpool L3 3AF, England, UK}
\altaffiltext{4}{Department of Physics and Astronomy, Dearborn Observatory, Northwestern University, 2131 Sheridon Road, Evanston, IL 60208}

\begin{abstract}

From the Press-Schechter mass function and the empirical X-ray cluster luminosity-temperature ($L$-$T$) relation, we construct an X-ray cluster luminosity function that can be applied to the growing number of high-redshift, X-ray cluster luminosity catalogs to constrain cosmological parameters.  
In this paper, we apply this luminosity function to the {\it Einstein} Medium Sensitivity Survey (EMSS) and the {\it ROSAT} Brightest Cluster Sample (BCS) luminosity function to constrain the value of $\Omega_m$.  
In the case of the EMSS, we find a factor of 4 $-$ 5 fewer X-ray clusters at redshifts above $z = 0.4$ than below this redshift at luminosities above $L_X = 7 \times 10^{44}$ erg s$^{-1}$ (0.3 - 3.5 keV), which suggests that the X-ray cluster luminosity function has evolved above $L_\star$.  At lower luminosities, this luminosity function evolves only minimally, if at all.
Using Bayesian inference, we find that the degree of evolution at high luminosities suggests that $\Omega_m = 0.96^{+0.36}_{-0.32}$, given the best-fit $L$-$T$ relation of Reichart, Castander, \& Nichol (1998).  When we account for the uncertainty in how the empirical $L$-$T$ relation evolves with redshift, we find that $\Omega_m \approx 1.0 \pm 0.4$.  However, it is unclear to what degree systematic effects may affect this and similarly obtained results.

\end{abstract}


\section{Introduction}

The combination of the Press-Schechter mass function (e.g., Press \& Schechter 1974; Oukbir \& Blanchard 1992; Lacey \& Cole 1993) and present and future X-ray cluster catalogs presents a unique opportunity to constrain the cosmological mass density parameter, $\Omega_m$.  The Press-Schechter approach offers a number of advantages over various, more traditional methods of measuring this parameter.  First of all, numerical simulations reproduce the Press-Schechter mass function to a high degree of accuracy (e.g., Eke \etal 1996; Bryan \& Norman 1998; Borgani \etal 1998).  
Secondly, unlike methods that only probe $\Omega_m$ over small spatial scales $-$ methods that may be insensitive to an underlying, more uniformly distributed component of the dark matter $-$ the Press-Schechter approach probes $\Omega_m$ over the scales of X-ray cluster catalogs, which now have limiting redshifts of about unity.  
Thirdly, the Press-Schechter approach appears to be relatively insensitive to a cosmological constant (e.g., Henry 1997; Mathiesen \& Evrard 1998; Eke \etal 1998; Viana \& Liddle 1998); consequently, a Press-Schechter-based determination of $\Omega_m$ might be compared to independent determinations of the deceleration parameter, for example, to constrain the cosmological constant.  Finally and most importantly, a number of independent, high-redshift, X-ray cluster luminosity catalogs with well-understood selection functions are and will soon be available (see \S4).  
We discuss potential problems with the Press-Schechter approach in \S4.
	
The archetypal X-ray cluster catalog is the X-ray cluster subsample of the {\it Einstein} Medium Sensitivity Survey, which we refer to here as the EMSS.  A complete description of this sample and its selection criteria can be found in Henry \etal (1992) (see also Gioia \etal 1990b; Stocke \etal 1991; Gioia \& Luppino 1994).  The original EMSS consists of 93 X-ray clusters, of which 67 have redshifts $z \ge 0.14$.  Nichol \etal (1997) updated this $z \ge 0.14$ subsample with 21 {\it ROSAT} luminosities and optical information from the literature.  This revised EMSS consists of 64 X-ray clusters, of which 25 have been updated.  The redshift and luminosity ranges of the revised EMSS are $0.140 \le z \le 0.823$ and $7.5 \times 10^{43} \le L_X \le 2.336 \times 10^{45}$ erg s$^{-1}$ (0.3 - 3.5 keV).
The EMSS is of great importance because at present, it is the only X-ray cluster catalog that probes masses above $M_\star$ at high redshifts, where the Press-Schechter mass function is the most sensitive to $\Omega_m$ (see \S3).  

Assuming that X-ray clusters correspond to virialized, dark matter halos, the Press-Schechter mass function describes how the X-ray-selected cluster mass function evolves with redshift.  Unfortunately, X-ray-selected cluster mass catalogs that span sufficiently broad ranges in $M$ and $z$ to constrain $\Omega_m$ do not yet exist.
Since the Press-Schechter mass function already assumes that X-ray clusters are virialized, one may convert this mass function to a temperature function with the virial theorem (see \S2); however, X-ray cluster temperature catalogs that span sufficiently broad ranges in $T$ and $z$ to strongly constrain $\Omega_m$ also do not yet exist (Viana \& Liddle 1998; Blanchard, Bartlett, \& Sadat 1998; however, see Henry 1997; Eke \etal 1998).  However, several X-ray cluster luminosity catalogs span sufficiently broad ranges in $L$ and $z$ to strongly constrain $\Omega_m$, and the number of such catalogs is growing.  
However, to fit the Press-Schechter mass function to such catalogs, one must invoke a luminosity-temperature ($L$-$T$) relation in addition to the virial theorem.  Theoretically, a wide variety of $L$-$T$ relations have been proposed (e.g., Kaiser 1986; Evrard \& Henry 1991; Kaiser 1991); consequently, the $L$-$T$ relation should be determined empirically.  Until recently, the $L$-$T$ relations of temperature catalogs have suffered from much scatter (e.g., Edge \& Stewart 1991; David \etal 1993; Mushotzky \& Scharf 1997); however, recently Markevitch (1998), Allen \& Fabian (1998), and Arnaud \& Evrard (1998) have published temperature catalogs with temperatures and luminosities that have either been corrected for, or avoided the effects of cooling flows; the result is a significant reduction of this scatter.  
Using Bayesian inference, Reichart, Castander, \& Nichol (1998) have constrained the slope and the evolution of the empirical $L$-$T$ relation for the luminosity range $10^{44.5}$ erg s$^{-1}$ $\la$ $L_{bol} \la 10^{46.5}$ erg s$^{-1}$ and the redshift range $z \la 0.5$ from the Markevitch (1998) and Allen \& Fabian (1998) catalogs.

This latter work may be the key to determining cosmological parameters with X-ray cluster catalogs:  given a well-constrained $L$-$T$ relation, the Press-Schechter mass function may be fitted to the growing number of independent, high-redshift, high-luminosity X-ray cluster catalogs with well-understood selection functions.
Using the cooling flow corrected $L$-$T$ relation of Reichart, Castander, \& Nichol (1998), we do this to the EMSS and the {\it ROSAT} Brightest Cluster Sample (BCS) luminosity function of Ebeling \etal (1997) in \S3.  In \S2, we model the X-ray cluster luminosity function; in \S4, we draw conclusions and discuss future applications of this luminosity function to the Southern SHARC and the Bright SHARC.

\section{The Model}

\subsection{The X-ray Cluster Luminosity Function}

Assuming that X-ray clusters correspond to virialized, dark matter halos, we model the comoving number density of X-ray clusters with the Press-Schechter mass function, which is given by (e.g., Lacey \& Cole 1993) 
\begin{equation}
\frac{dn_c(M,z)}{dM} = -\sqrt{\frac{2}{\pi}} \frac{\bar{\rho}_0}{M^2} \frac{d\ln{\sigma_0}(M)}{d\ln{M}} \frac{\delta_{c0}(z)}{\sigma_0(M)} \exp{\left[\frac{-\delta_{c0}^2(z)}{2\sigma_0^2(M)}\right]},
\label{ps}
\end{equation}
where $n_c(M,z)$ is the comoving number density of X-ray clusters of mass $M$ at redshift $z$, $\delta_{c0}(z)$ is the present linear theory overdensity of perturbations that collapsed and virialized at redshift $z$, $\sigma_0(M)$ is the present linear theory variance of the mass density fluctuation power spectrum filtered on mass scale $M$, and $\bar{\rho}_0$ is the present mean mass density of the universe.  In the case of zero cosmological constant, which we assume throughout this paper, the present overdensity is given by (Lacey \& Cole 1993; Peebles 1980) 
\begin{equation}
\delta_{c0}(z) = \cases{\frac{3}{2} D(0) \left(\left(\frac{2\pi}{\sinh{\eta}-\eta}\right)^{2/3}+1\right) & ($\Omega_m < 1$) \cr \frac{3}{20} (12\pi)^{2/3} (1+z) & ($\Omega_m = 1$) \cr \frac{3}{2} D(0) \left(\left(\frac{2\pi}{\eta-\sin{\eta}}\right)^{2/3}-1\right) & ($\Omega_m > 1$)},
\label{del}
\end{equation}  
where
\begin{equation}
D(0) = \cases{1 + \frac{3}{x_0} + \frac{3\sqrt{1+x_0}}{x_0^{3/2}} \ln{(\sqrt{1+x_0}-\sqrt{x_0})} & ($\Omega_m < 1$) \cr -1 + \frac{3}{x_0} - \frac{3\sqrt{1+x_0}}{x_0^{3/2}} \tan^{-1}{\sqrt{\frac{x_0}{1-x_0}}} & ($\Omega_m > 1$)},
\end{equation}
\begin{equation}
x_0 = \left|\Omega_m^{-1} - 1\right|,
\end{equation}
\begin{equation}
\eta = \cases{\cosh^{-1}{\left(\frac{2}{\Omega(z)} - 1\right)} & ($\Omega_m < 1$) \cr \cos^{-1}{\left(1 - \frac{2}{\Omega(z)}\right)} & ($\Omega_m > 1$)},
\end{equation}
and
\begin{equation}
\Omega(z) = \frac{\Omega_m (1+z)}{1 + \Omega_m z}.
\end{equation}
We assume a scale-free mass density fluctuation power spectrum of power law index $n$, so the present variance is given by (e.g., Lacey \& Cole 1993)
\begin{equation}
\sigma_0(M) = \sigma_8 \left(\frac{M}{M_8}\right)^{-\frac{3+n}{6}},
\end{equation}
where $\sigma_8$ is the amplitude of the mass density fluctuation power spectrum over spheres of radius $8 h^{-1}$ Mpc, and $M_8$ is the mean mass within these spheres.

We now convert equation (\ref{ps}) from a mass function to an appropriately defined luminosity function.  Following the notation of Mathiesen \& Evrard (1998), we begin by assuming that X-ray clusters' bolometric luminosities scale as power laws in mass and redshift:
\begin{equation}
L_{bol} \propto M^p (1+z)^s. 
\label{power}
\end{equation}
As did Henry \etal (1992) in the case of the EMSS, we find that the fractions of this luminosity that fall into the EMSS band of 0.3 - 3.5 keV and the BCS band of 0.1 - 2.4 keV are well approximated by power laws in X-ray cluster temperature:
\begin{equation}
L_X = f_X T^{-\beta} L_{bol},
\label{fjc}
\end{equation}
where $f_X = 0.989 \pm 0.014$ and $\beta = 0.407 \pm 0.008$ for the representative temperature range of the EMSS ($3 \la T \la 10$ keV), and $f_X = 1.033 \pm 0.012$ and $\beta = 0.472 \pm 0.008$ for the representative temperature range of the BCS ($1.5 \la T \la 12$ keV), where temperature is measured in keV.  At lower temperatures, however, these approximations quickly fail.  Equation (\ref{fjc}) is independent of redshift because X-ray cluster luminosities are measured in the source frame.  The temperature dependence introduced by equation (\ref{fjc}) is removed with the virial theorem:
\begin{equation}
T \propto M^\frac{2}{3} (1+z).
\label{virial}
\end{equation}
Technically, this expression holds only when $\Omega_m = 1$; however, we show in \S4 that generalizing this expression has little effect upon our results.
Together, equations (\ref{power}), (\ref{fjc}), and (\ref{virial}) yield the following expression that relates an X-ray cluster's mass to its observed luminosity, $L_X$:
\begin{equation}
L_X \propto f_XM^{p-\frac{2\beta}{3}}(1+z)^{s-\beta}.
\label{factor}
\end{equation}
Substitution of equation (\ref{factor}) into equation (\ref{ps}) yields the following luminosity function:
\begin{equation}
\frac{dn_c(L_X,z)}{dL_X} = f(z)L_X^{-1-\frac{3-n}{2(3p-2\beta)}} \exp{\left[-g(z)L_X^{\frac{3+n}{3p-2\beta}}\right]},
\label{ps2}
\end{equation}
where
\begin{equation}
f(z) = af_X^{\frac{3-n}{2(3p-2\beta)}}(1+z)^{\frac{(s-\beta)(3-n)}{2(3p-2\beta)}}\delta_{c0}(z),
\end{equation}
\begin{equation}
g(z) = cf_X^{-\frac{3+n}{3p-2\beta)}}(1+z)^{-\frac{(s-\beta)(3+n)}{3p-2\beta}} \delta_{c0}^2(z),
\end{equation}
and $a$ and $c$ depend upon $\sigma_8$ and the factor of proportionality of equation (\ref{factor}).  Instead of trying to model this factor of proportionality and fitting to $\sigma_8$, we simply fit to such degenerate, or grouped, combinations of these parameters in this paper (see \S2.2).

Since luminosities are computed from measured fluxes and redshifts, $L_X$ is a function of $H_0$ and $\Omega_m$.  With one exception, all dependences upon $H_0$ can be grouped into the parameters $a$ and $c$, and this exception is noted in \S2.2.  The EMSS and the BCS provide luminosities in their respective X-ray bands that have been computed for $H_0 = 50$ km s$^{-1}$ Mpc$^{-1}$ and $\Omega_m = 1$; we denote these luminosities by $L_1$.  The relationship between $L_X$ and $L_1$ is given by 
\begin{equation}
L_X = x(z)L_1,
\label{lxl}
\end{equation}  
where 
\begin{equation}
x(z) = \cases{\left(\frac{d_L(\Omega_m)}{d_L(\Omega_m = 1)}\right)^2 \frac{f_F(d_A(\Omega_m = 1))}{f_F(d_A(\Omega_m))} & ({\it Einstein}) \cr \left(\frac{d_L(\Omega_m)}{d_L(\Omega_m = 1))}\right)^2 & ({\it ROSAT})},
\label{x}
\end{equation}
$d_L(\Omega_m)$ is luminosity distance, $d_A(\Omega_m)$ is angular diameter distance, and $f_F(d_A(\Omega_m))$ is the fraction of an X-ray cluster's flux that is detected in the 2\arcmin.4 $\times$ 2\arcmin.4 detect cell of the original EMSS.\footnote{The quantity $f_F(d_A(\Omega_m))$ is also a function of X-ray cluster core radius, $a_0$, which Henry \etal (1992) found to be $a_0 \sim 0.25$ Mpc, assuming that $\Omega_m = 1$.  Repeating their analysis for $\Omega_m = 0$, we find that this result again holds; consequently, we adopt this value of $a_0$ throughout this paper.}  A complete description of this quantity can be found in Henry \etal (1992).  Since {\it ROSAT} measures total fluxes, $f_F(d_A(\Omega_m)) = 1$ here.  
So in the case of the BCS, the latter expression applies.  However, the revised EMSS subsample that we fit to in \S3.3 is a combination of 43 {\it Einstein} luminosities and 18 {\it ROSAT} luminosities (see \S3).  Fortunately, 36 of the 43 {\it Einstein} clusters have redshifts of $z < 0.33$, and 6 of the remaining 7 clusters have redshifts of $z < 0.47$.  
At these redshifts, the ratio of fractional fluxes in the former expression for $x(z)$ is within a few percent of unity for a wide range of values of $\Omega_m$.  Furthermore, {\it ROSAT} luminosities are available for 7 of the 8 clusters that carry the majority of the weight in the fits of \S3.3, and the remaining {\it Einstein} cluster is at a redshift of $z = 0.259$.  Consequently, we also use the latter expression for $x(z)$ in the case of the EMSS.  
Besides, we show in \S3 that the sensitivity of $x(z)$ to $\Omega_m$ plays only a tertiary role in the determination of this parameter.  
Substitution of equation (\ref{lxl}) into equation (\ref{ps2}) allows the luminosity function to be fitted to $L_1$ data without loss of generality:
\begin{equation}
\frac{dn_c(L_1,z)}{dL_1} = f(z)L_1^{-1-\frac{3-n}{2(3p-2\beta)}} \exp{\left[-g(z
)L_1^{\frac{3+n}{3p-2\beta}}\right]},
\label{ps3}
\end{equation}
where
\begin{equation}
f(z) = af_X^{\frac{3-n}{2(3p-2\beta)}}(1+z)^{\frac{(s-\beta)(3-n)}{2(3p-2\beta)}} \delta_{c0}(z) x^{-\frac{3-n
}{2(3p-2\beta)}}(z),
\label{f}
\end{equation}
and 
\begin{equation}
g(z) = cf_X^{-\frac{3+n}{3p-2\beta}}(1+z)^{-\frac{(s-\beta)(3+n)}{3p-2\beta}} \delta_{c0}^2(z) x^{\frac{3+n}
{3p-2\beta}}(z).
\label{g}
\end{equation}

\subsection{The Selection Function}

Let $A(L_1,z)$ be the area of the sky that an X-ray survey samples at redshift $z$ as a function of luminosity $L_1$.  In the case of the EMSS, this quantity is given by (Avni \& Bahcall 1980; Henry \etal 1992; Nichol \etal 1997)
\begin{equation}
A(L_1,z) = A(F_{lim} = F(L_1,z)),
\label{area}
\end{equation}
where $A(F_{lim})$ is the area of the sky that the EMSS surveyed below sensitivity limit $F_{lim}$ (see Henry \etal 1992), 
\begin{equation}
F(L_1,z) = \frac{f_F(d_A(z))}{k(z)} \frac{h_{50}^2L_1}{4\pi d_L^2(z)},
\label{flux}
\end{equation}
and $k(z)$ is the k-correction from the observer frame to the source frame for a $T = 6$ keV X-ray cluster; the exact dependence of $k(z)$ upon X-ray cluster temperature can be ignored for the representative temperature range of the EMSS.
For the EMSS, we have computed $A(L_1,z)$ for 41 values of $L_1$ between $10^{43.5}$ and $10^{45.5}$ erg s$^{-1}$ (0.3 - 3.5 keV) and for $\Omega_m = 0$, 0.5, 1, and 1.5.  For intermediate values of $L_1$ and $\Omega_m$, we use linear interpolation between $43.5 < \log{L_1} < 45.5$ and $0 < \Omega_m < 1.5$.
The cases of $\Omega_m = 0$ and 1 are plotted in Figure 1.  
The dependence of $A(L_1,z)$ upon $H_0$ cannot be grouped into the parameters $a$ and $c$, unlike all of the other $H_0$ dependences in this analysis (\S2.1).  Instead of making $H_0$ a free parameter, we fix $H_0 = 50$ km s$^{-1}$ Mpc$^{-1}$ in this paper.  However, if others wish to be more general, they need only consider the $H_0$ dependence of the selection function.
The case of the BCS is treated separately in \S3.2.

The total number of X-ray clusters observed between luminosity and redshift limits $L_l < L_1 < L_u$ and $z_l < z < z_u$, i.e., the cumulative luminosity function, is given by
\begin{equation}
N(L_l,L_u;z_l,z_u) = \int_{L_l}^{L_u}\int_{z_l}^{z_u}A(L_1,z)\frac{dn_c(L_1,z)}{dL_1}dL_1dV(z),
\label{ps4}
\end{equation}
where 
\begin{equation}
dV(z) = \frac{4c^3dz}{H_0^3\Omega_m^4(1+z)^3}\frac{(\Omega_m z + (\Omega_m-2)((\Omega_m z+1)^\frac{1}{2}-1))^2}{(1+\Omega_m z)^\frac{1}{2}}  
\label{vol}
\end{equation}
is the comoving volume element.
Hence, our model (luminosity function $+$ selection function) consists of nine parameters:  $H_0$, $f_X$, $\beta$, $p$, $s$, $a$, $c$, $n$, and $\Omega_m$.  We have fixed the value of $H_0$, and we have tightly constrained the values of $f_X$ and $\beta$ for the EMSS and the BCS (\S2.1).  In \S3.1, we adopt values of $p$ and $s$ from the Bayesian inference analysis of the X-ray cluster $L$-$T$ relation of Reichart, Castander, \& Nichol (1998), and in \S3.3, the normalization parameter, $a$, drops out of the Bayesian inference analysis of this paper.  This leaves three parameters:  $\Omega_m$, $n$, and the grouped parameter $c$, which depends upon $\sigma_8$, the proportionality constant of equation (\ref{factor}), and $H_0$.  We determine credible intervals for the values of these three parameters in \S3.3.

\section{Bayesian Inference}

If properly applied, equation (\ref{ps4}) can be an effective probe of $\Omega_m$.  In this cumulative luminosity function, $\delta_{c0}(z)$, $A(L_1,z)$, $x(z)$, and $dV(z)$ depend upon $\Omega_m$.  We now consider how sensitive each of these quantities is to $\Omega_m$.  The comoving volume element $dV(z)$, is approximately given by (equation (\ref{vol}))
\begin{equation}
\frac{dV(z)}{dz} \propto \cases{\frac{z^{2.26}}{(1+z)^3} & ($\Omega_m = 0$) \cr \frac{z^{1.99}}{(1+z)^3} & ($\Omega_m = 1$)},
\end{equation}
where the indices apply in the redshift range $0.14 < z < 0.6$.  
The luminosity conversion expression, $x(z)$, is approximately given by (equation (\ref{x}))
\begin{equation}
x(z) \approx 1 - \frac{1-\Omega_m}{4}z,
\end{equation}
and in equation (\ref{ps3}), it is always raised to a power that is between $\approx$ $-1.5$ and $\approx$ 0.5.  Consequently, $dV(z)$ and $x(z)$ contribute only weak dependences upon $\Omega_m$ to equation (\ref{ps4}).  The present overdensity is a stronger function of $\Omega_m$ (equation (\ref{del})):
\begin{equation}
\delta_{c0}(z) = \cases{1.5 & ($\Omega_m = 0$) \cr 1.69(1+z) & ($\Omega_m = 1$)}.
\label{del2}
\end{equation}
Since this expression appears to the second power in the exponential cutoff of equation (\ref{ps3}), it contributes a significant dependence upon $\Omega_m$ to the cumulative luminosity function.  For example, if the luminosity function is observed to cut off prematurely at higher redshifts, i.e., if there is a deficit of high-redshift, luminous X-ray clusters, then higher values of $\Omega_m$ are favored.  However, if little or no evolution is manifest in the observed X-ray cluster luminosity function, particularly above $L_\star$\footnote{In this paper, $L_\star$ refers very generally to those luminosities at which the luminosity function, modeled by equation (\ref{ps3}), appears to roll over from a power law to an exponential cutoff.}, then lower values of $\Omega_m$ are favored.  

The surveyed area, $A(L_1,z)$, contributes a different type of dependence upon $\Omega_m$ to the cumulative luminosity function.  In the case of the EMSS (see Figure 1), this dependence is negligible at low luminosities and redshifts.  However, at luminosities $\ga L_\star$, $A(L_1,z)$ is a non-negligible, increasing function of $\Omega_m$ at sufficiently high redshifts.  
For example, in the case of a $L_1 = 10^{45}$ erg s$^{-1}$, z = 0.8 EMSS cluster, $A(L_1,z)$ is roughly twice as large in an $\Omega_m = 1$ universe than it is in an $\Omega_m = 0$ universe.  Although this effect is suppressed by the exponential cutoff of equation (\ref{ps3}) above $L_\star$, this effect is amplified about $L_\star$ by the fact that the luminosity function itself is a non-negligible, increasing function of $\Omega_m$ at luminosities $\la L_\star$ at these high redshifts (see \S3.3).  
Consequently, we find that an overabundance of high-redshift, $\sim L_\star$ EMSS clusters favors higher values of $\Omega_m$ and not lower values of this parameter as is generally thought.  We return to this idea in \S3.3.

Consider first the case of X-ray cluster luminosity data that lies within a narrow redshift band of effective redshift $z_{eff}$.  Then, up to a factor of $A(L_1,z_{eff})dV(z_{eff})/dz$, the integrand of equation (\ref{ps4}) (equation (\ref{ps3})) is simply a power law in luminosity with an $\Omega_m$-dependent exponential cutoff.  
This exponential cutoff is a function of the parameters $f_X$, $\beta$, $p$, $n$, and $g_{eff} = g(z_{eff})$, which itself is a function of $\Omega_m$ (see below).  We have already constrained the values of $f_X$ and $\beta$ (\S2.1), and we adopt the Reichart, Castander, \& Nichol (1998) value of $p$, as well as that of $s$, in \S3.1.  However, there are too few high-luminosity X-ray clusters to simultaneously constrain $n$ and $g_{eff}$.  Fortunately, $n$ is also constrained by the low-luminosity, power-law limit of equation (\ref{ps3}) for which data is more plentiful. 
Consequently, by fitting this luminosity function to data of this type, $n$ and $g_{eff}$ can be jointly constrained.  

By equation (\ref{g}), the parameter $g_{eff}$ is a function of $z_{eff}$ and the parameters $f_X$, $\beta$, $p$, $s$, $n$, $c$, and $\Omega_m$.  The effective redshift is a given and the parameters $f_X$, $\beta$, $p$, $s$, and $n$ can be constrained as described above.  
However, since $z \approx z_{eff}$, a constant, the parameters $c$ and $\Omega_m$ are degenerate; consequently, $\Omega_m$ can only be constrained if the value of $c$ is otherwise known, i.e., if the values of $\sigma_8$, the factor of proportionality of equation (\ref{factor}), and $H_0$ are otherwise known (\S2.2).  
Even in the event that X-ray cluster mass data is used instead of luminosity data, any fitted value of $\Omega_m$ will still depend strongly upon the assumed value of $\sigma_8$, as well as upon the assumed value of $n$, since mass data is not yet plentiful enough for the Press-Schechter mass function to constrain these parameters.

However, now consider X-ray cluster luminosity data that spans a breadth of redshifts.  Instead of constraining the single parameter $g_{eff}$, one instead constrains a distribution of such parameters with redshift, i.e., $g(z)$.  The normalization of this distribution is $c$ and its shape yields $\Omega_m$ since the parameters $f_X$, $\beta$, $p$, $s$, and $n$ are otherwise constrained.  Consequently, by fitting equation (\ref{ps4}) to the EMSS, which spans a breadth of luminosities and redshifts, the parameters $n$, $c$, and $\Omega_m$ can be jointly constrained regardless of the values of $\sigma_8$ and the factor of proportionality of equation (\ref{factor}) (but not regardless of the value of $H_0$ since $A(L_1,z)$ is a function of this parameter (\S2.2)).  We do this for the EMSS in \S3.3.
In \S3.2, we better constrain the parameters $n$ and $c$ (actually, $g_{eff}$) with the local ($z_{eff} \sim 0.1$) luminosity function of the BCS.  First however, we discuss the cooling flow corrected $L$-$T$ relation of Reichart, Castander, \& Nichol (1998) and its implied values of $p$ and $s$, in \S3.1.

\subsection{The $L$-$T$ Relation}

The combination of equations (\ref{power}) and (\ref{virial}) yields the $L$-$T$ relation:
\begin{equation}
L_{bol} \propto T^\frac{3p}{2}(1+z)^{s-\frac{3p}{2}}.
\label{lt}
\end{equation}
Reichart, Castander, \& Nichol (1998) have constrained the slope and the evolution of equation (\ref{lt}) using the cooling flow corrected X-ray cluster temperature catalogs of Markevitch (1998) and Allen \& Fabian (1998), and Bayesian inference.  For the luminosity ranges of the EMSS and the BCS, and the redshift range $z \la 0.5$, they find that $p = 1.86^{+0.10}_{-0.10}$ and $s = (3.77 - 0.63\Omega_m)^{+0.48}_{-1.22}$.  However, when using the $L$-$T$ relation to fit the Press-Schechter mass function to X-ray cluster luminosity catalogs (1) that are not cooling flow corrected, and (2) for which X-ray cluster photon count rates have been converted to fluxes and luminosities by assuming a $T = 6$ keV thermal bremsstrahlung spectrum, they find that one should use $p = 1.77^{+0.16}_{-0.13}$ and $s = (3.14 - 0.65\Omega_m)^{+0.88}_{-0.86}$.  

In the case of the EMSS, this latter case applies.  In this paper, we adopt the best-fit values:  $p = 1.77$ and $s = 3.14 - 0.65\Omega_m$.  The parameter $p$ is well-constrained, and in \S3.3, we show that the uncertainty in the value of $s$ does not significantly affect our results.  
In the case of the BCS, luminosities are not cooling flow corrected, but they are not determined by assuming $T = 6$ keV for each X-ray cluster; instead, luminosities are determined by additionally requiring that each X-ray cluster satisfy a $L$-$T$ relation (Ebeling \etal 1997).  
From Figure 1 of Reichart, Castander, \& Nichol (1998), it is apparent that the assumption of $T = 6$ keV more strongly affects the value of $p$ than does the use of cooling flow corrected luminosities.  Consequently, in this paper, we adopt $p = 1.86$ in the case of the BCS.  We note, however, that such minor variations in this parameter do not significantly affect our results.  
Since the BCS is a local ($z_{eff} \sim 0.1$) catalog, the value of $s$ is unimportant in this case (see \S3.2).

\subsection{The {\it ROSAT} BCS}

The {\it ROSAT} BCS is a flux-limited sample of 199 bright X-ray clusters.  A complete description of this sample and its selection criteria can be found in Ebeling \etal (1997).  The redshift range of the BCS is $z < 0.3$; however, most of the BCS clusters have redshifts of $z < 0.2$, and the effective redshift (see \S3) of the sample is $z_{eff} \sim 0.1$.  
Consequently, the BCS samples the X-ray cluster population of the local universe.  Although such a sample may not have enough redshift leverage to adequately probe $\Omega_m$, its large size makes it an excellent sample to constrain the parameters $n$ and $c$ (actually, $g_{eff}$, \S3).  These constraints can then be combined with the EMSS results of \S3.3 to better constrain $\Omega_m$.  However, since the BCS is not yet publicly available, we settle here for a simplified analysis of the binned BCS luminosity function of Figure 1 of Ebeling \etal (1997), which we replot in Figure 2.

Since the BCS spans a relatively narrow band of redshifts, we let $f(z) = f(z_{eff}) = f_{eff}$ and $g(z) = g(z_{eff}) = g_{eff}$ in equation (\ref{ps3}).  This approximation is reasonable, unless the value of $\Omega_m$ is high, in which case one expects a lower comoving number density of X-ray clusters in the highest-luminosity bins.  This is because the highest-luminosity bins more strongly sample the highest-redshift BCS clusters than do the lower-luminosity bins.  Given this approximation $-$ that $z = z_{eff}$, a constant $-$ equation (\ref{ps3}) can only reproduce such a high-luminosity roll-over of the luminosity function by favoring an artificially high value of $n$.
To safeguard against this potential bias, we first fit equation (\ref{ps3}) to all 12 luminosity bins, then to all but the highest-luminosity bin, then to all but the 2 highest-luminosity bins, etc., until the fitted values of $n$ and $c$ do not change appreciably from fit to fit.  
Also, since Ebeling \etal (1997) have already corrected this luminosity function for sample completeness, we set $A(L_1,z_{eff}) = 1$.  Here, we ignore the dependence that this quantity has upon $\Omega_m$, which is a reasonable approximation since $z_{eff} < 0.3$.

We find that only the highest-luminosity bin noticeably changes our results.  When we fit equation (\ref{ps3}) to all 12 luminosity bins, we find that $n = -0.47^{+0.32}_{-0.31}$ and $g_{eff} = 0.90^{+0.28}_{-0.14}$, where we have assumed a flat prior probability distribution between $-3 < n < 0$ and $0 < g_{eff} < 10$, and the likelihood function is given by $e^{-\chi^2/2}$.\footnote{See, e.g., Gregory \& Loredo (1992) for an excellent discussion of Bayesian inference.}  When we fit equation (\ref{ps3}) to all but the highest-luminosity bin, we find that $n = -1.83^{+0.85}_{-0.15}$ and $g_{eff} = 1.20^{+0.70}_{-0.60}$.  Ignoring additional high-luminosity bins does not appreciably change this result; consequently, in this paper, we use all but the highest-luminosity bin to determine a constraint from the BCS.  We will explore what this highest-luminosity bin implies for the value of $\Omega_m$ in a later paper.  
In Figure 3, we plot the 1, 2, and 3 $\sigma$ credible regions in the $n-g_{eff}$ plane for both of the above fits.  We do not plot credible regions in the $n-c$ plane because, by equation (\ref{g}), the parameter $c$ is a degenerate function of the parameters $g_{eff}$ and $\Omega_m$ (see \S3); however, credible regions in the $n-c$ plane are easily recovered, given a value of $\Omega_m$ (see \S3.3).
In Figure 2, we also plot the best-fit luminosity functions of these fits.

\subsection{The EMSS}

As described in \S3, the breadth of the luminosity and redshift ranges of the EMSS makes this catalog an ideal sample with which to probe $\Omega_m$.
In Figure 4, we plot the $L_1-z$ distribution of the revised EMSS of Nichol \etal (1997), as well as the $z < 0.14$ portion of the original EMSS X-ray cluster subsample.  The solid curves are contours of constant sampled differential volume, i.e., $A(L_1,z)dV(z)/dz =$ constant.  From left to right, these contours are equally spaced from zero (zero contour not shown).  
If the X-ray cluster luminosity function has not evolved over the redshift range of the EMSS, then at each luminosity, most of the observed X-ray clusters would be where most of the sampled differential volume is.  This appears to be the case below $L_1 \sim 7 \times 10^{44}$ erg s$^{-1}$ (0.3 - 3.5 keV), which demonstrates agreement with the results of Collins \etal (1997), Nichol \etal (1997), Burke \etal (1997), Rosati \etal (1998), Jones \etal (1998), and Vikhlinin \etal (1998a), i.e., that the X-ray cluster luminosity function evolves only minimally, if at all, below $L_\star$ (see \S4).  
However, as was originally found by Gioia \etal (1990a) (see also Henry \etal 1992), there appears to be a deficit of high-redshift X-ray clusters above $L_\star$.  For example, given that six X-ray clusters were detected between $0.14 < z < 0.4$ at $L_1 > 7 \times 10^{44}$ erg s$^{-1}$ (0.3 - 3.5 keV), one would expect $\sim 9$ X-ray clusters between $0.4 < z < 0.6$ and $\sim 21$ X-ray clusters between $0.4 < z < 0.9$ at these luminosities if the luminosity function were not evolving, yet only 2 and 4 clusters were detected in these redshift ranges, respectively.  Hence, we find a factor of 4 $-$ 5 fewer X-ray clusters at redshifts above $z = 0.4$ than below this redshift at these luminosities.  This is only possible at the $\approx 1 - 2$\% level with the no evolution model.  
This suggests that high values of $\Omega_m$ may be favored (\S3); however, we offer alternative interpretations of this deficit in \S4.

By Bayes' theorem, the posterior probability distribution for $\Omega_m$, $n$, and $c$, $P(\Omega_m,n,c)$, is given by normalizing the product of the prior probability distribution and the likelihood function (e.g., Gregory \& Loredo 1992).  Here, we assume a flat prior probability distribution between $0 < \Omega_m < 1.5$, $-3 < n < 0$, and $0 < c < 3$.  The likelihood function, ${\cal{L}}(\Omega_m,n,c)$, is given by (e.g., Cash 1979) 
\begin{equation}
{\cal{L}}(\Omega_m,n,c) = \prod_{i=1}^{N_{tot}}P(L_{1,i},z_i|\Omega_m,n,c),
\label{likely}
\end{equation}
where $P(L_{1,i},z_i|\Omega_m,n,c)$ is the probability that the $i$th X-ray cluster fits our model, given values of $\Omega_m$, $n$, and $c$.  For our model (equation (\ref{ps4})), this probability is given by (e.g., Cash 1979)
\begin{equation}
P(L_1,z|\Omega_m,n,c) = \frac{A(L_1,z)}{N(L_l,L_u;z_l,z_u)} \frac{dV(z)}{dz} \frac{dn_c(L_1,z)}{dL_1}.
\end{equation}
Since both $dn_c(L_1,z)/dL_1$ and $N(L_l,L_u;z_l,z_u)$ are proportional to the normalization parameter $a$, our results are clearly independent of this parameter.

In equation (\ref{likely}), $N_{tot}$ is the total number of X-ray clusters in the same region of the $L_1-z$ plane as that over which $N(L_l,L_u;z_l,z_u)$ is defined.  This region should be as broad as is reasonably possible and it need not be rectangular, as the simple integration limits of equation (\ref{ps4}) suggest.  
In this paper, we set $L_l < L_1 < L_u  = 10^{45.5}$ erg s$^{-1}$, where $L_l$ is set by the limiting flux of the EMSS:  $F(L_l,z) = 1.33$ x 10$^{-13}$ erg cm$^{-2}$ s$^{-1}$ (Henry \etal 1992); hence, $L_l$ is a function of redshift.\footnote{In computing $L_l(z)$, we set $\Omega_m = 0$.  This is because if we were to use a higher value of $\Omega_m$ when defining this curve, the selection function, $A(L_1,z)$, would be undefined for luminosities and redshifts near this curve when the value of $\Omega_m$ is lower than the curve-defining value (see equation (\ref{flux})).}
Also in this paper, we set $0.14 = z_l < z < z_u = 0.6$.  We exclude higher redshifts because (1) the $L$-$T$ relation of Reichart, Castander, and Nichol (1998) is derived from $z \la 0.5$ X-ray clusters, so its accuracy should not be trusted at redshifts much in excess of this value, and (2) optical identification become more difficult at these high redshifts (\S1).  This excludes three $\sim$ $L_\star$ X-ray clusters, which reduces the total number of X-ray clusters in our sample from 64 to 61.
Below, however, we repeat our analysis with $z_u = 0.9$ to show that the exclusion of these three X-ray clusters does not unfairly bias our results.
In Figure 4, we mark this region of the $L_1-z$ plane with a dotted line; solid points are interior to this region.

The posterior probability distribution for any two parameters, e.g., $P(\Omega_m,n)$, or one parameter, e.g., $P(\Omega_m)$, is given by marginalizing the posterior probability distribution for all three parameters, $P(\Omega_m,n,c)$, over the other parameters (e.g., Gregory \& Loredo 1992).  
1, 2, and 3 $\sigma$ credible regions are determined by integrating the posterior probability distribution over the most probable region of its parameter space until 68.3\%, 95.4\% and 99.73\% of this distribution has been integrated (e.g., Gregory \& Loredo 1992).
In Figure 5, we plot the 1, 2, and 3 $\sigma$ credible regions of the two-dimensional posterior probability distributions $P(\Omega_m,n)$ (top panel) and $P(n,c)$ (bottom panel).  The 1 $\sigma$ credible intervals of the one-dimensional posterior probability distributions, i.e., $P(\Omega_m)$, $P(n)$, and $P(c)$, are $\Omega_m = 0.96^{+0.48}_{-0.38}$, $n = -2.28^{+0.36}_{-0.25}$, and $c = 0.66^{+0.48}_{-0.23}$.
Hence, the EMSS favors high values of $\Omega_m$ and low values of $n$.
Although these results by themselves are constraining $-$ $\Omega_m < 0.2$ is ruled out at the 2.3 $\sigma$ credible level $-$ by combining the likelihood function of the EMSS with the posterior probability distribution of the BCS (Figure 3), stronger constraints can be placed upon these parameters.

We now determine the combined posterior probability distribution of the EMSS and the BCS.  Let the likelihood function be that of the EMSS, as given by equation (\ref{likely}).  However, instead of assuming a flat prior probability distribution for all three parameters, as we did above, only assume a flat prior probability distribution between $0 < \Omega_m < 1.5$, and use the posterior probability distribution of the BCS $-$ $P_{BCS}(n,g_{eff})$ (Figure 3) $-$ equation (\ref{g}), and the effective redshift of the BCS, $z_{eff} \sim 0.1$, to determine the full prior probability distribution:  $P_{BCS}(\Omega_m,n,c)$.
In Figure 6, we plot credible regions of the two-dimensional combined EMSS/BCS posterior probability distributions $P(\Omega_m,n)$ and $P(n,c)$.  In the three left panels of Figure 8, we plot the one-dimensional combined EMSS/BCS posterior probability distributions.  The dotted lines in this figure mark the 1, 2, and 3 $\sigma$ credible intervals.  We find that $\Omega_m = 0.96^{+0.36}_{-0.32}$, $n = -1.86^{+0.42}_{-0.34}$, and $c = 0.54^{+0.24}_{-0.12}$.  Values of $\Omega_m < 0.2$ are ruled out at the 3.0 $\sigma$ credible level.  

To establish that the exclusion of the three $z > 0.6$, $\sim$ $L_\star$ EMSS clusters does not unfairly bias our results, we repeat this analysis for $z_u = 0.9$.  In Figure 7, we plot credible regions of the two-dimensional combined EMSS/BCS posterior probability distributions $P(\Omega_m,n)$ and $P(n,c)$.  In the three right panels of Figure 8, we plot the one-dimensional combined EMSS/BCS posterior probability distributions.  In this case, we find that $\Omega_m = 0.93^{+0.33}_{-0.26}$, $n = -1.50^{+0.37}_{-0.36}$, and $c = 0.48^{+0.12}_{-0.12}$.  Values of $\Omega_m < 0.2$ are ruled out at the 3.5 $\sigma$ credible level.
Lower values of $\Omega_m$ are not favored for three reasons.  First of all, it is clear from Figure 3 that we have added a great deal of volume for which there are no EMSS clusters above $L_\star$.  
Secondly, for the region of the $L_1-z$ plane occupied by these three clusters, the EMSS surveyed roughly twice as much area if $\Omega_m = 1$ than it did if $\Omega_m = 0$ (\S3).  This is primarily because higher values of $\Omega_m$ imply lower luminosity distances, which imply higher fluxes for a given luminosity and, consequently, greater surveyed areas by equation (\ref{area}).  The third reason, also mentioned in \S3, is that at luminosities $\la L_\star$, the luminosity function itself is a non-negligible, increasing function of $\Omega_m$.  At these luminosities, the redshift dependence of the luminosity function is dominated by the function $f(z)$.  This function is approximately given by (equations (\ref{f}), (\ref{del}), and (\ref{x}))
\begin{equation}
f(z) \propto \cases{(1+z)^{0.33(3-n)} & ($\Omega_m = 0$) \cr (1+z)^{1+0.23(3-n)} & ($\Omega_m = 1$)}
\end{equation}
Consequently, there is an additional factor of $\approx (1+z)^{1/2}$ in the $\Omega_m = 1$ case.  
Although our $z_u = 0.9$ results are more constraining than are our $z_u = 0.6$ results, for the reasons stated above, we feel less confident about these results than we do about our $z_u = 0.6$ results.

We now show that the uncertainty in the value of $s$ does not significantly affect our results.  In the three left panels of Figure 9, we plot the one-dimensional combined EMSS/BCS posterior probability distributions, once again for $z_u = 0.6$, except that we have now used the $-1$ $\sigma$ value of $s$ from Reichart, Castander, \& Nichol (1998).  In the three right panels of Figure 9, we plot the same distributions, but for the $+1$ $\sigma$ value of $s$ from Reichart, Castander, \& Nichol (1998).  
The effect of varying the value of $s$ by $\pm 1$ $\sigma$ is a variation in the values of the fitted parameters by less than the extent of their $\pm 1$ $\sigma$ uncertainties.  If we add these uncertainties in quadrature, we find that $\Omega_m \approx 1.0 \pm 0.4$, $n \approx -1.9 \pm 0.4$, and $c \approx 0.55^{+0.25}_{-0.15}$.  One-dimensional credible intervals for all of the fitted values in this section are compiled in Table 1.

\section{Discussion \& Conclusions}

In this paper, we have constructed from the Press-Schechter mass function and the empirical X-ray cluster $L$-$T$ relation of Reichart, Castander, \& Nichol (1998) an X-ray cluster luminosity function that can be applied to the growing number of independent, high-redshift, X-ray cluster luminosity catalogs to constrain cosmological parameters.  
In particular, we have incorporated the evolution of the $L$-$T$ relation and all significant dependences upon $\Omega_m$ of the luminosity and selection functions into our Bayesian inference analysis.  For a fixed value of $H_0$, we have applied this luminosity function to broad subsets of the revised EMSS X-ray cluster subsample of Nichol \etal (1997) and to the {\it ROSAT} BCS luminosity function of Ebeling \etal (1997) to constrain $\Omega_m$.  For the 61 revised EMSS clusters between $0.14 < z < 0.6$, we find that $\Omega_m = 0.96^{+0.36}_{-0.32}$ and $n = -1.86^{+0.42}_{-0.34}$; for all 64 revised EMSS clusters, between $0.14 < z < 0.9$, we find that $\Omega_m = 0.93^{+0.33}_{-0.26}$ and $n = -1.50^{+0.37}_{-0.36}$.
These high values of $\Omega_m$ are the result of an apparent deficit of high-redshift, luminous X-ray clusters, which suggests that the X-ray cluster luminosity function has evolved above $L_\star$.  

Nichol \etal (1997) suggested that the statistical evidence for the evolution of the EMSS luminosity function was only minimal.  At first glance, this appears to be in contradiction to one of the conclusions of this paper.  However, since Nichol \etal (1997) used a power-law luminosity function - which they did for purposes of comparison with the original EMSS result of Henry \etal (1992) - instead of a luminosity function that permits different degrees of evolution below and above $L_\star$, as we have done in this paper, their results are most directly applicable below $L_\star$:  this is the luminosity range of the vast majority of the EMSS clusters, so it is by this luminosity range that their results have most strongly been weighted.
The fact that the X-ray cluster luminosity function does not evolve below $L_\star$ has since been shown by Collins \etal (1997), Burke \etal (1997), Rosati \etal (1998), Jones \etal (1998), and Vikhlinin \etal (1998a).  
That the luminosity function appears to evolve above $L_\star$ is in agreement with the original EMSS findings of Gioia \etal (1990a), as well as the findings of Vikhlinin \etal (1998a,b) with the 160 deg$^2$ survey. 

The value of $\Omega_m$ that we find that this high-luminosity evolution in the EMSS corresponds to is consistent with the values found by Sadat, Blanchard, \& Oukbir (1998) ($\Omega_m = 0.85 \pm 0.2$) and Blanchard \& Bartlett (1998) ($\Omega_m \approx 1$), based upon the work of Oukbir \& Blanchard (1992,1997).  Our value of $\Omega_m$ is somewhat consistent with the values found by Henry (1997) ($\Omega_m = 0.50 \pm 0.14$) and Eke \etal (1998) ($\Omega_m = 0.45 \pm 0.2$); however, Viana \& Liddle (1998) have performed a more extensive error analysis upon a conservative subset of the data of these authors and find that $\Omega_m \sim 0.75$ with $\Omega_m > 0.3$ at the 90\% confidence level and $\Omega_m \sim 1$ still viable.
Blanchard, Bartlett, \& Sadat (1998) find almost identical results ($\Omega_m \sim 0.74$, with $0.3 < \Omega_m < 1.2$ at the 95\% confidence level) from these data.
Finally, our value of $\Omega_m$ is inconsistent with the values found by Bahcall, Fan, \& Cen (1997) ($\Omega_m = 0.3 \pm 0.1$), Fan, Bahcall, \& Cen (1997) ($\Omega_m \approx 0.3 \pm 0.1$), and Bahcall \& Fan (1998) ($\Omega_m = 0.2^{+0.3}_{-0.1}$).  

Our value of $n$ is consistent with the values found by Henry \& Arnaud (1991) ($n = -1.7^{+0.65}_{-0.35}$) and Henry \etal (1992) ($n = -2.10^{+0.27}_{-0.15}$), where these authors set $\Omega_m = 1$.  Our value of $n$ is also consistent with the value found by Eke \etal (1998) ($n = -1.69^{+0.12}_{-0.07}$), where these authors included $\Omega_m$ as a free parameter.  Our value of $n$ is somewhat consistent with the value that Bahcall, Fan, \& Cen (1997), Fan, Bahcall, \& Cen (1997), and Bahcall \& Fan (1998) adopted ($n = -1.4$).

Taken as an ensemble, these results are perhaps discouraging in that they span the entire range of acceptable solutions:  $0.2 \la \Omega_m \la 1$.  This suggests that as yet unknown systematic effects may be plaguing some, if not all, of these results.  We briefly identify seven areas where systematic effects could enter ours and similar analyses.
(1) The first is the Press-Schechter mass function itself; however, numerical simulations (e.g., Eke \etal 1996; Bryan \& Norman 1997; Borgani \etal 1998) consistently show that the Press-Schechter mass function is an adequate approximation.  
(2) The spherical collapse model of cluster formation (equations (\ref{del}) and (\ref{del2})) may be inadequate.  For example, numerical simulations by Governato \etal (1998) suggest that in equation (\ref{del2}), the expression $1.69(1+z)$ may really be as low as $\sim 1.6(1+z)^{0.9}$.  This suggests that use of the spherical collapse model may lead to underestimated values of $\Omega_m$; however, this is only a $\la 10$\% effect.
(3) Technically, equation (\ref{virial}) only holds when $\Omega_m = 1$.  Recently, Voit \& Donahue (1998) derived a virial theorem that holds for all values of $\Omega_m$, and that allows for the fact that clusters grow gradually; their mass-temperature ($M$-$T$) relation reduces to equation (\ref{virial}) when $\Omega_m = 1$.  We find that $M$-$T$ relations with functional forms that are similar to that of the $M$-$T$ relation of Voit \& Donahue (1998) reduce our fitted value of $\Omega_m$ by $\la 10$\%; however, further investigation and use of this $M$-$T$ relation is clearly needed.
(4) Also on the subject of the $M$-$T$ relation, care must be taken when fitting to X-ray cluster temperature catalogs:  cooling flows lower the measured temperature of most X-ray clusters, which should systematically affect values of $\Omega_m$ that are determined in this way.
(5) Based upon the cooling flow corrected X-ray cluster temperature catalogs of Markevitch (1998) and Allen \& Fabian (1998), Reichart, Castander, \& Nichol (1998) determined an empirical $L$-$T$ relation between measured luminosities and cooling flow corrected temperatures that holds for $z \la 0.5$ and for luminosities that are typical of X-ray cluster catalogs (see \S3.1); however, more cooling flow corrected X-ray cluster temperature measurements are needed to determine what, if any, exceptions exist to this $L$-$T$ relation, and to extend it to higher redshifts.
(6) The art of determining an X-ray cluster catalog's selection function is a constantly improving science; modern selection functions are determined via extensive numerical simulations.  An alternative explanation to our high-$\Omega_m$ result is that the EMSS, for whatever reasons, missed many high-redshift, high-luminosity X-ray clusters beyond what is accounted for by their selection function (see \S3.3).  However, given that the EMSS detected many high-redshift, low-luminosity X-ray clusters, this seems to be an unlikely scenario.
(7) Finally, our cosmological model may be inadequate.  We did not investigate the effects of a cosmological constant in this paper; however, many authors have demonstrated that the inclusion of a cosmological constant has little effect upon the determined value of $\Omega_m$ (see \S1).  Also, the effects of quintessent and other exotic cosmologies have not yet been investigated in this context.

Some of these potential sources of systematic error can be safeguarded against.  For example, the spherical collapse model, the $M$-$T$ relation, and the $L$-$T$ relation all have proportionality factors that are potential sources of systematic error.  However, as we have shown in \S2.1, all of these factors, as well as the parameter $\sigma_8$, group together, giving us our parameter $c$.  Since we fit for $c$, these factors cannot bias our result.  However, inadequate functional forms for these relations, as well as for the other functions listed above, can bias ours and others' results.

In addition to further theoretical and numerical development of this formalism, only the continued construction of X-ray cluster catalogs will act to further resolve these issues.  Fortunately, the number of X-ray cluster luminosity catalogs is growing rapidly.  One such catalog is the Southern Serendipitous High-Redshift Archival {\it ROSAT} Catalog (Southern SHARC) (Collins \etal 1997; Burke \etal 1997).  The redshift and luminosity ranges of the Southern SHARC are $z < 0.7$ and $L_1 < 3 \times 10^{44}$ erg s$^{-1}$ (0.5 - 2.0 keV).  Although the Southern SHARC does not span the luminosity range of the EMSS, it will provide a good consistency check of our EMSS results.  Our analysis of this catalog is underway.

Two similar X-ray cluster catalogs that can serve a similar purpose are the {\it ROSAT} Deep Cluster Survey (RDCS) (Rosati \etal 1998) and the Wide Angle {\it ROSAT} Pointed Survey (WARPS) (Jones \etal 1998).  The RDCS spans the redshift and luminosity ranges $z < 0.8$ and $L_1 < 3 \times 10^{44}$ erg s$^{-1}$ (0.5 - 2.0 keV).  The WARPS spans the redshift and luminosity ranges $z < 0.7$ and $L_1 < 2 \times 10^{44}$ erg s$^{-1}$ (0.5 - 2.0 keV).  An analysis of the RDCS is also underway (Borgani \etal 1998). 

The 160 deg$^2$ survey (Vikhlinin \etal 1998b) and the Bright SHARC (Romer \etal 1998), a high-luminosity extension of the Southern SHARC that is currently under construction, span redshift and luminosity ranges that rival those of the EMSS.  Consequently, these catalogs will provide strong, independent checks of the EMSS results.

Finally, local ($z_{eff} \sim 0.1$) X-ray cluster catalogs, such as the {\it ROSAT} BCS, are of great importance.  Although these samples do not have the redshift leverage to constrain cosmological parameters, their large sizes make them excellent samples to better constrain the parameters $n$ and $c$.  Samples like the EMSS, the 160 deg$^2$ survey, and the Bright SHARC do not have sufficient luminosity leverage to strongly constrain these parameters, which leads to weaker constraints upon the cosmological parameters.  However, a simultaneous analysis of a local X-ray cluster catalog $-$ as opposed to a local X-ray cluster luminosity function as we have used in this paper $-$ and any of these high-redshift, high-luminosity X-ray cluster catalogs could lead to significantly improved constraints upon all of these parameters.

\acknowledgments
This research has been partially funded by NASA grants NAG5-6548 and NAG5-2432.  We are very grateful to H. Ebeling for providing us with the data for Figure 2.  Also, we are grateful to A. Blanchard, S. Borgani, C. Graziani, D. Q. Lamb, C. A. Metzler, C. Scharf, M. S. Turner, and J. M. Quashnock for valuable discussions.  We are also very grateful to our anonymous referee, whose input has greatly improved this paper.  D. E. R. is especially grateful to Dr. and Mrs. Bernard Keisler for their hospitality during the summers of 1997 and 1998.  

\clearpage

\begin{deluxetable}{ccccccc}
\footnotesize
\tablecolumns{5}
\tablewidth{0pc}
\tablecaption{Fitted Values of the X-ray Cluster Luminosity Function}
\tablehead{\colhead{Catalog(s)} & \colhead{$z_u$} & \colhead{$s$\tablenotemark{a}} & \colhead{$\Omega_m$} & \colhead{$n$} & \colhead{$c$} & \colhead{$P(\Omega_m > 0.2$)}}
\startdata
BCS\tablenotemark{b} & $-$ & $-$ & $-$ & $-1.83^{+0.85}_{-0.15}$ & $-$ & $-$ \nl
EMSS & $0.6$ & $3.14-0.65\Omega_m$ & $0.96^{+0.48}_{-0.38}$ & $-2.28^{+0.36}_{-0.25}$ & $0.66^{+0.48}_{-0.23}$ & 2.3 $\sigma$ \nl
EMSS+BCS & $0.6$ & $3.14-0.65\Omega_m$ & $0.96^{+0.36}_{-0.32}$ & $-1.86^{+0.42}_{-0.34}$ & $0.54^{+0.24}_{-0.12}$ & 3.0 $\sigma$ \nl
EMSS+BCS & $0.9$ & $3.14-0.65\Omega_m$ & $0.93^{+0.33}_{-0.26}$ & $-1.50^{+0.37}_{-0.36}$ & $0.48^{+0.12}_{-0.12}$ & 3.5 $\sigma$ \nl
EMSS+BCS & $0.6$ & $2.26-0.61\Omega_m$ & $0.69^{+0.36}_{-0.27}$ & $-1.74^{+0.42}_{-0.36}$ & $0.48^{+0.28}_{-0.12}$ & 2.6 $\sigma$ \nl
EMSS+BCS & $0.6$ & $3.79-0.59\Omega_m$ & $1.17^{+0.33}_{-0.28}$ & $-1.98^{+0.34}_{-0.30}$ & $0.66^{+0.25}_{-0.18}$ & 3.2 $\sigma$ \nl
\enddata
\tablenotetext{a}{Reichart, Castander, \& Nichol 1998.}
\tablenotetext{b}{$g_{eff} = 1.20^{+0.70}_{-0.60}$.}
\end{deluxetable}

\clearpage

\figcaption[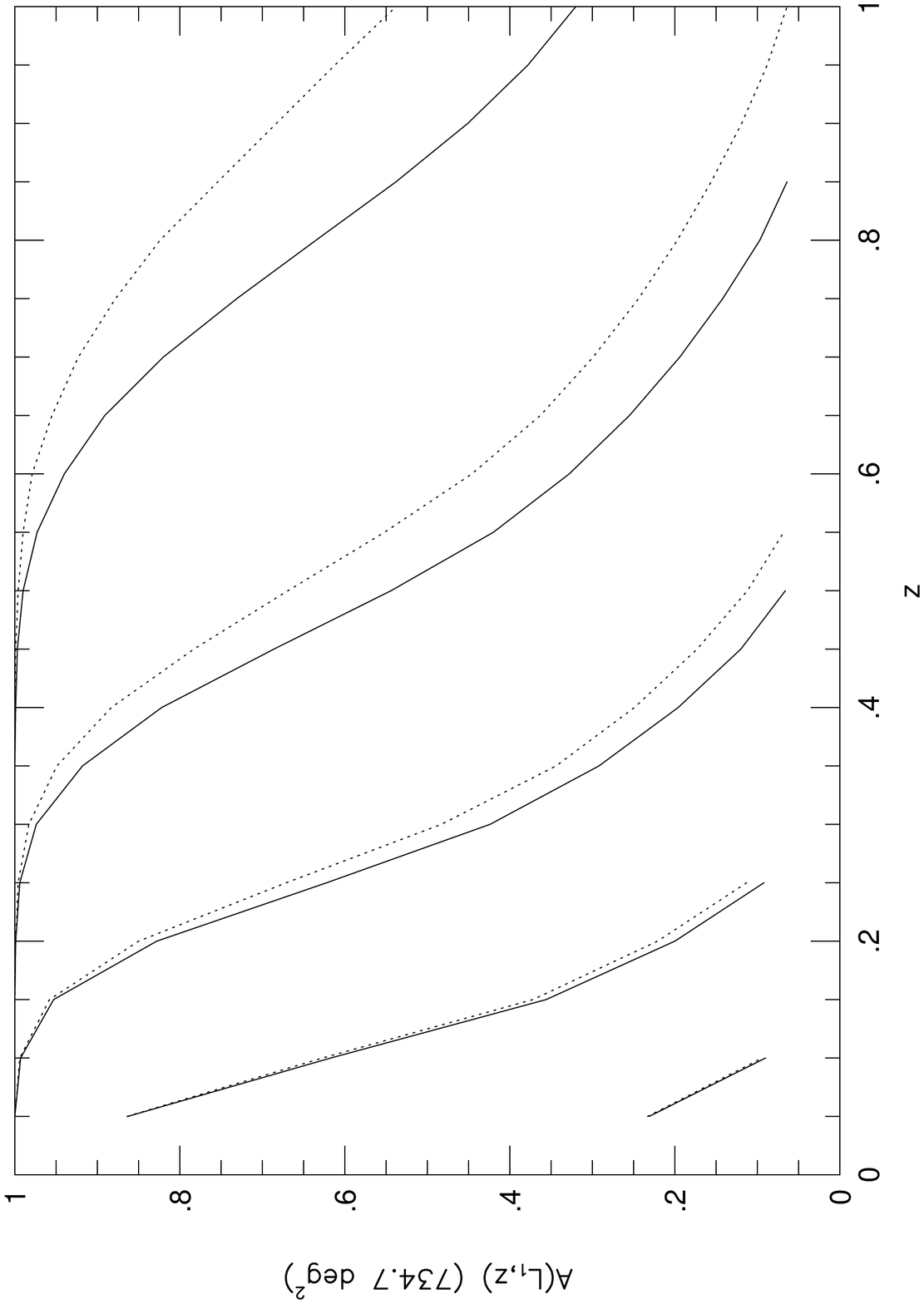]{The area $A(L_1,z)$ of the sky that the EMSS sampled at redshift $z$ as a function of luminosity $L_1$.  The solid curve is for $\Omega_m = 0$ and the dotted curve is for $\Omega_m = 1$.  From left to right, the curves correspond to $L_1 = 10^{43.5}$, $10^{44}$, $10^{44.5}$, $10^{45}$, and $10^{45.5}$ erg s$^{-1}$ (0.3 - 3.5 keV) (see \S2.2).\label{omfig1.ps}}

\figcaption[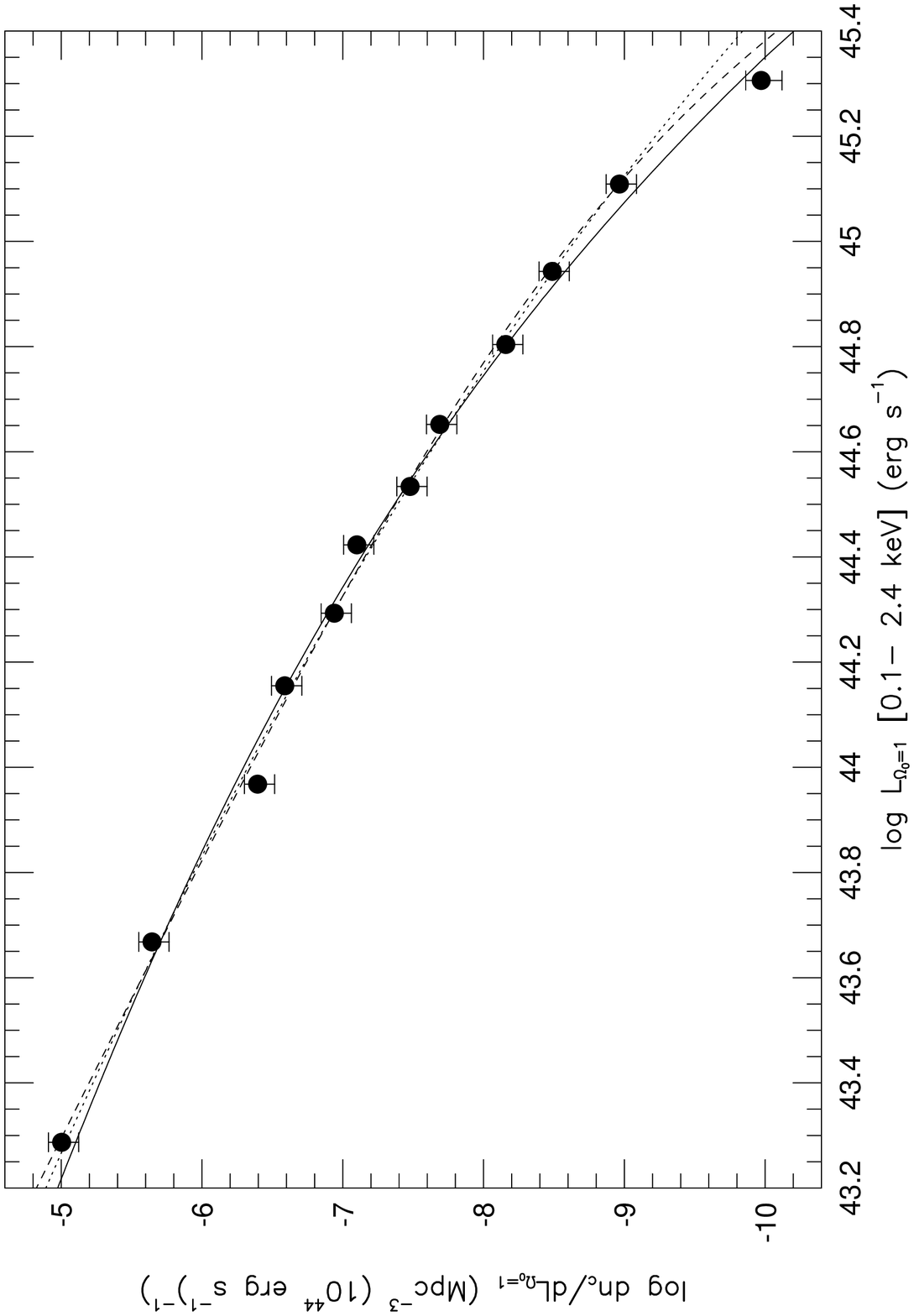]{The {\it ROSAT} BCS luminosity function of Ebeling \etal (1997).  The solid line is the best fit of equation (18) to all 12 luminosity bins.  The dotted line is the best fit of equation (18) to all but the highest-luminosity bin (see \S3.2).  The dashed line is the best-fit Schechter function of Ebeling \etal (1997).\label{omfig2.ps}}

\figcaption[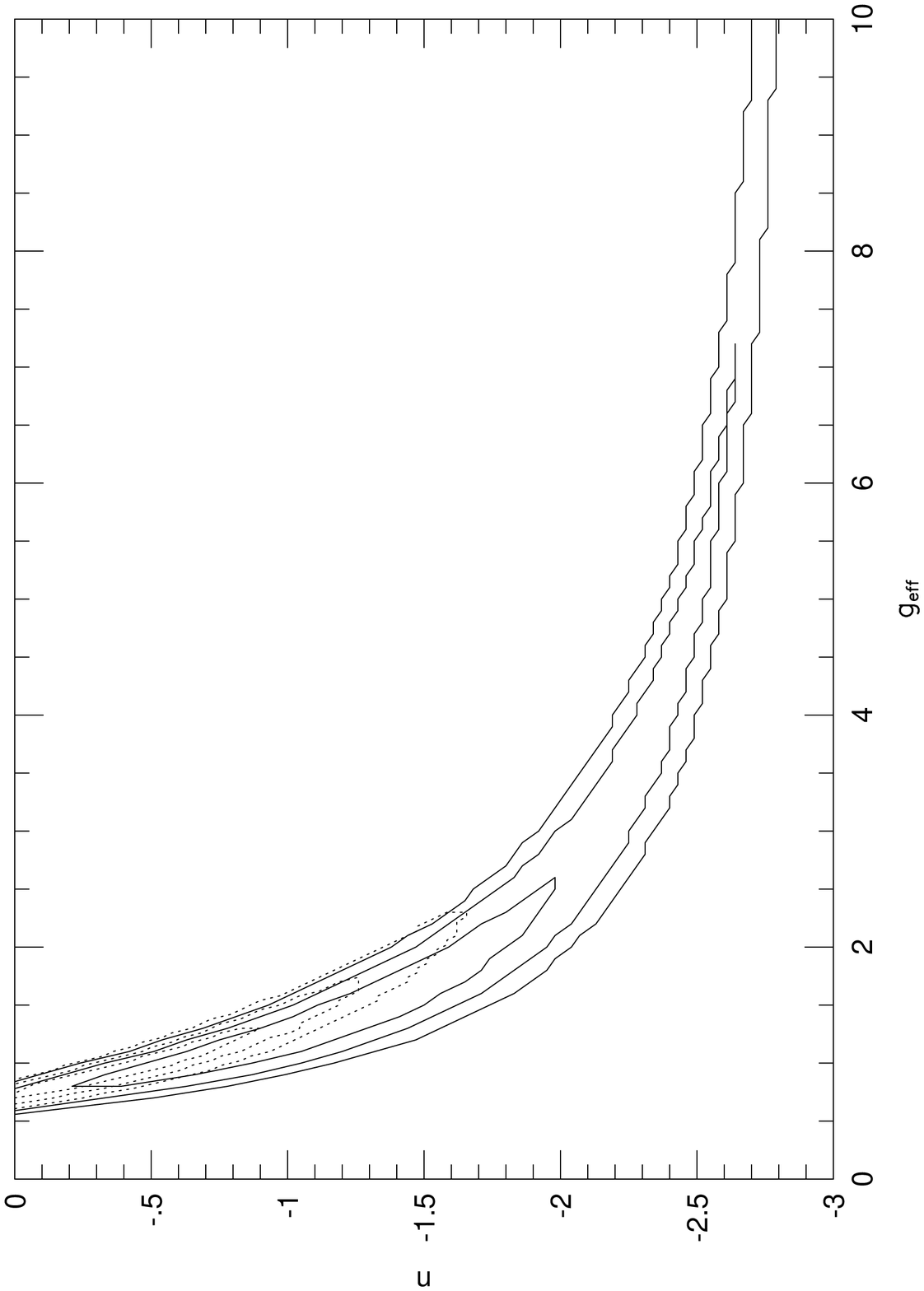]{The 1, 2, and 3 $\sigma$ credible regions of the posterior probability distributions of the fit of equation (18) to all 12 luminosity bins (dotted lines) and all but the highest-luminosity bin (solid lines) of the {\it ROSAT} BCS luminosity function (see \S3.2).\label{omfig3.ps}}

\figcaption[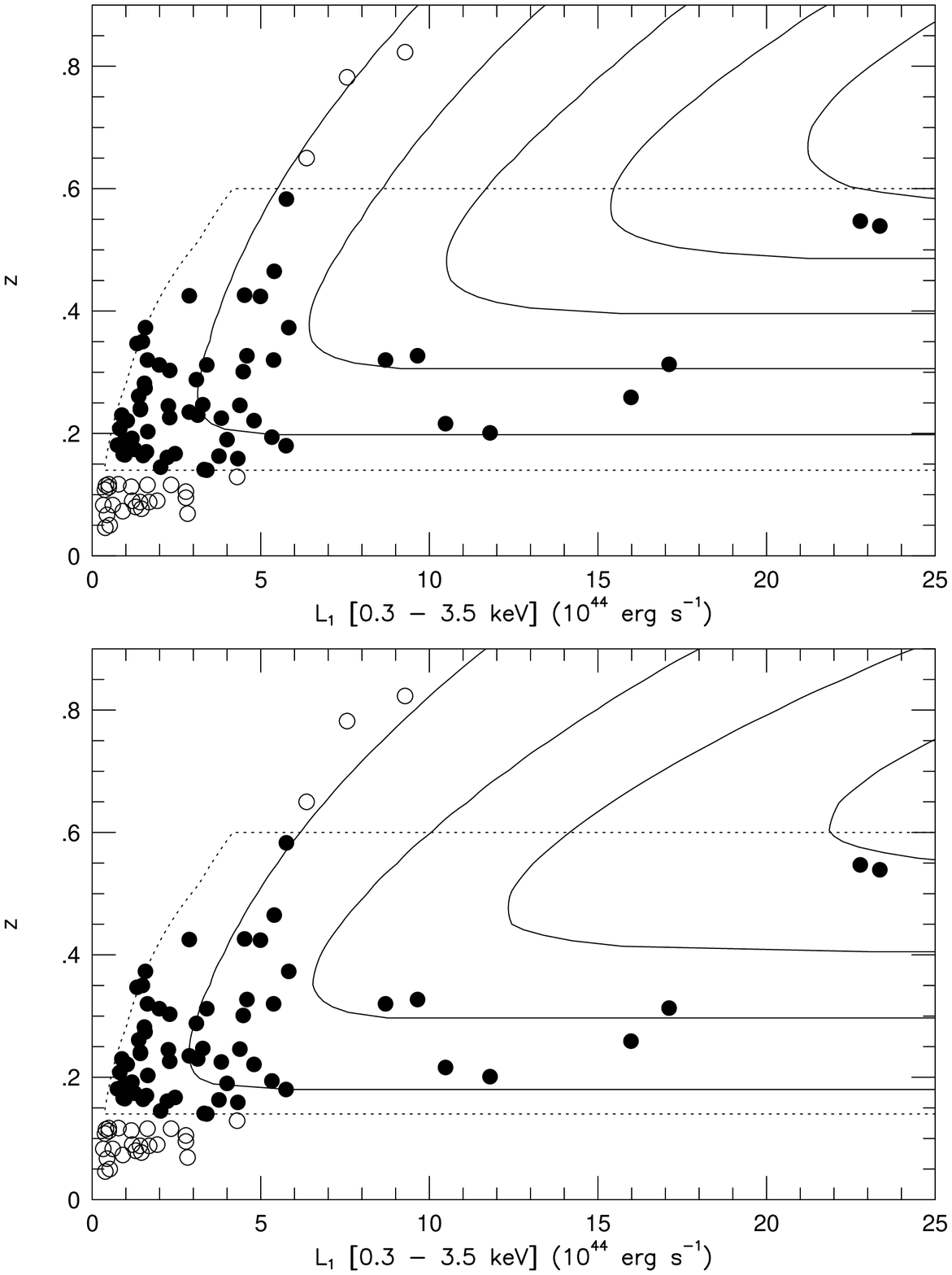]{The $L_1-z$ distribution of the EMSS clusters.  From left to right, the solid curves are increasing, equally spaced (from zero, zero contour not shown) contours of constant sampled differential volume (see \S3.3).  A deficit of high-redshift X-ray clusters is apparent above $L_1 \sim 7 \times 10^{44}$ erg s$^{-1}$.  The dotted curve is the more conservative of the two regions over which we fit equation (23) in \S3.3.  Points interior to this region are solid.  The top panel is for $\Omega_m = 0$ and the bottom panel is for $\Omega_m = 1$.\label{omfig4.ps}}

\figcaption[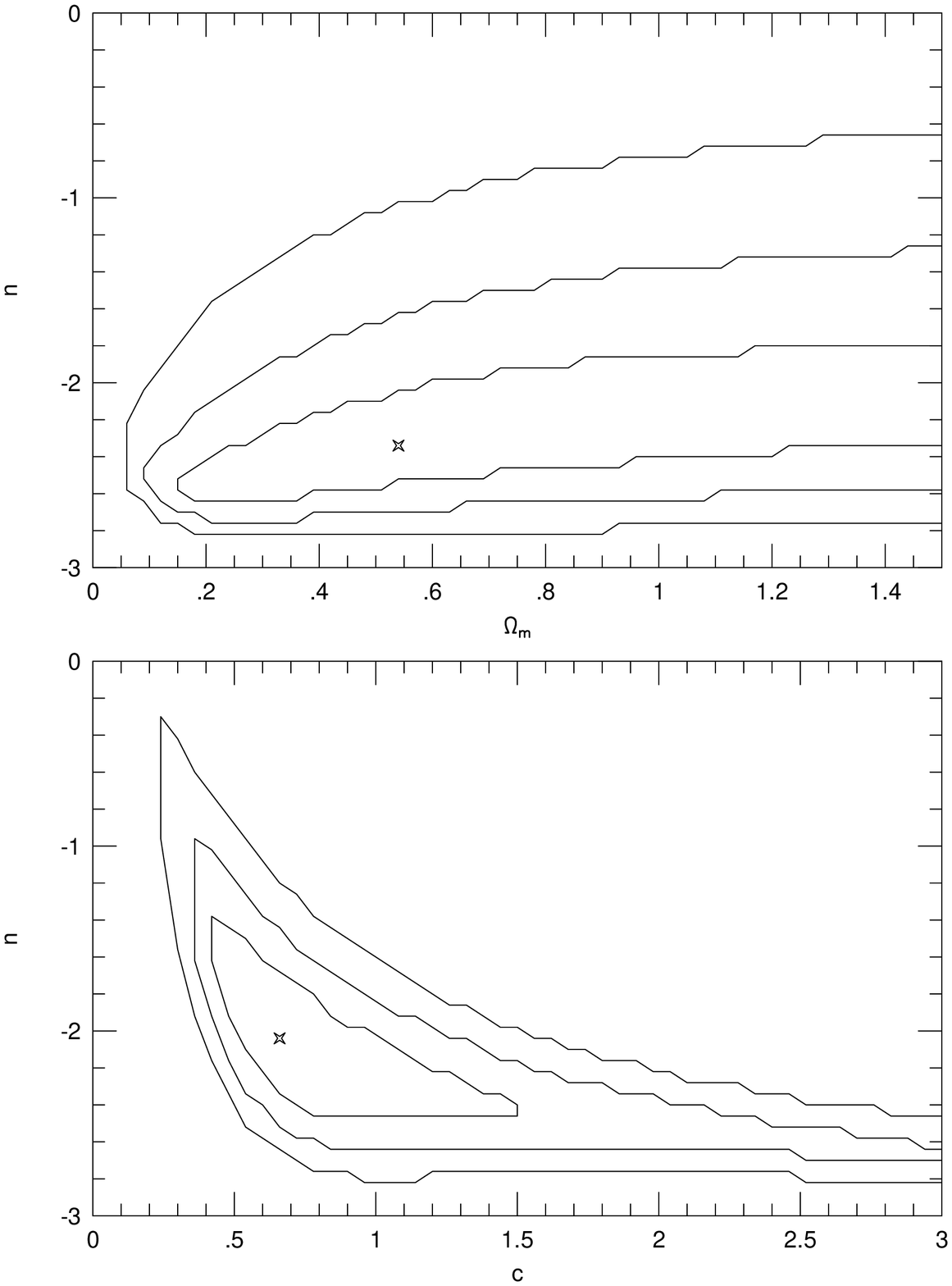]{The 1, 2, and 3 $\sigma$ credible regions of the marginalized posterior probability distributions $P(\Omega_m,n)$ (top panel) and $P(n,c)$ (bottom panel) of the fit of equation (23) to the $0.14 < z < 0.6$ revised EMSS clusters (see \S3.3).\label{omfig5.ps}}

\figcaption[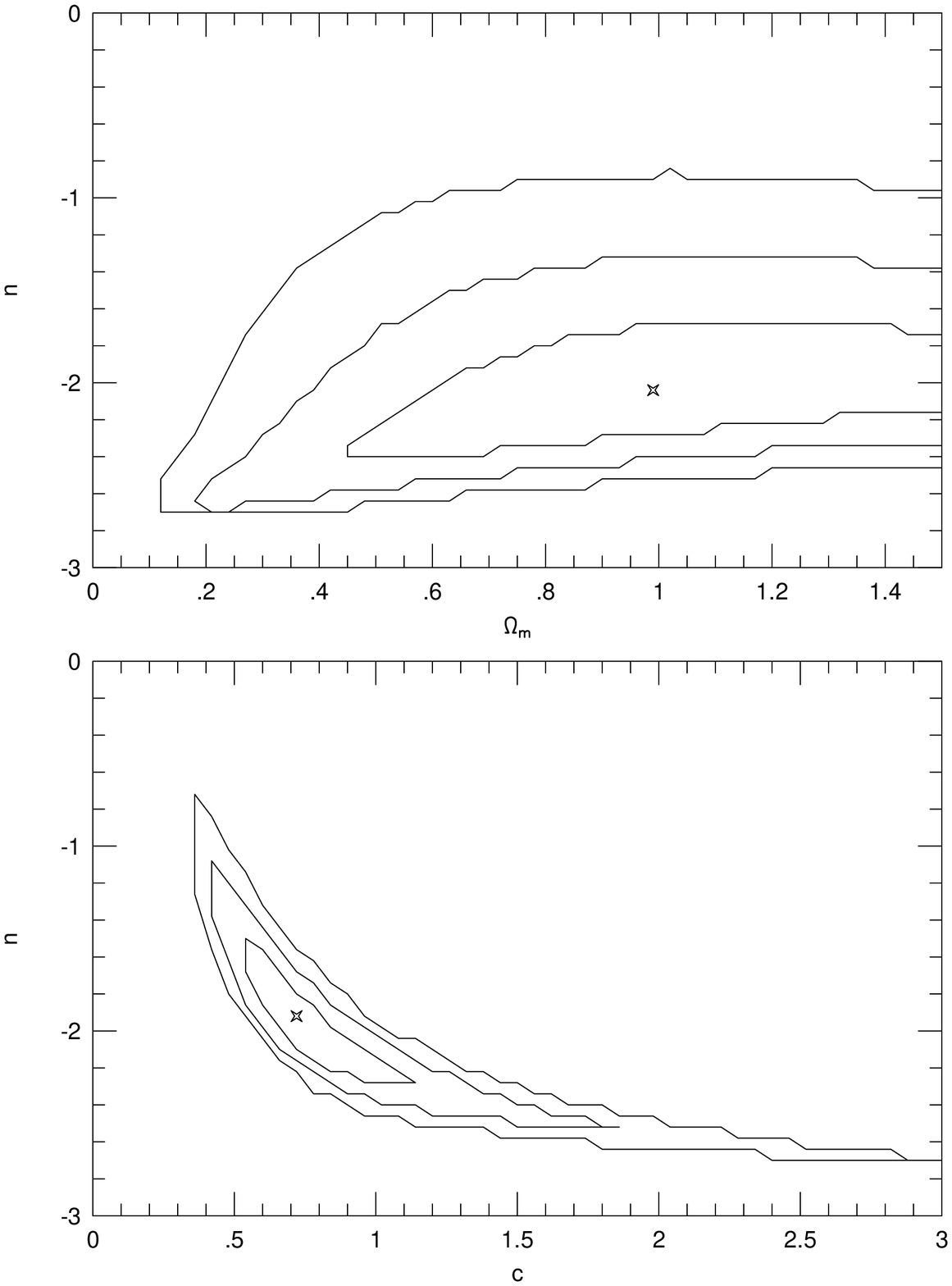]{The 1, 2, and 3 $\sigma$ credible regions of the marginalized posterior probability distributions $P(\Omega_m,n)$ (top panel) and $P(n,c)$ (bottom panel) of the fit of equation (23) to the $0.14 < z < 0.6$ revised EMSS clusters and the {\it ROSAT} BCS luminosity function (see \S3.3).\label{omfig6.ps}}

\figcaption[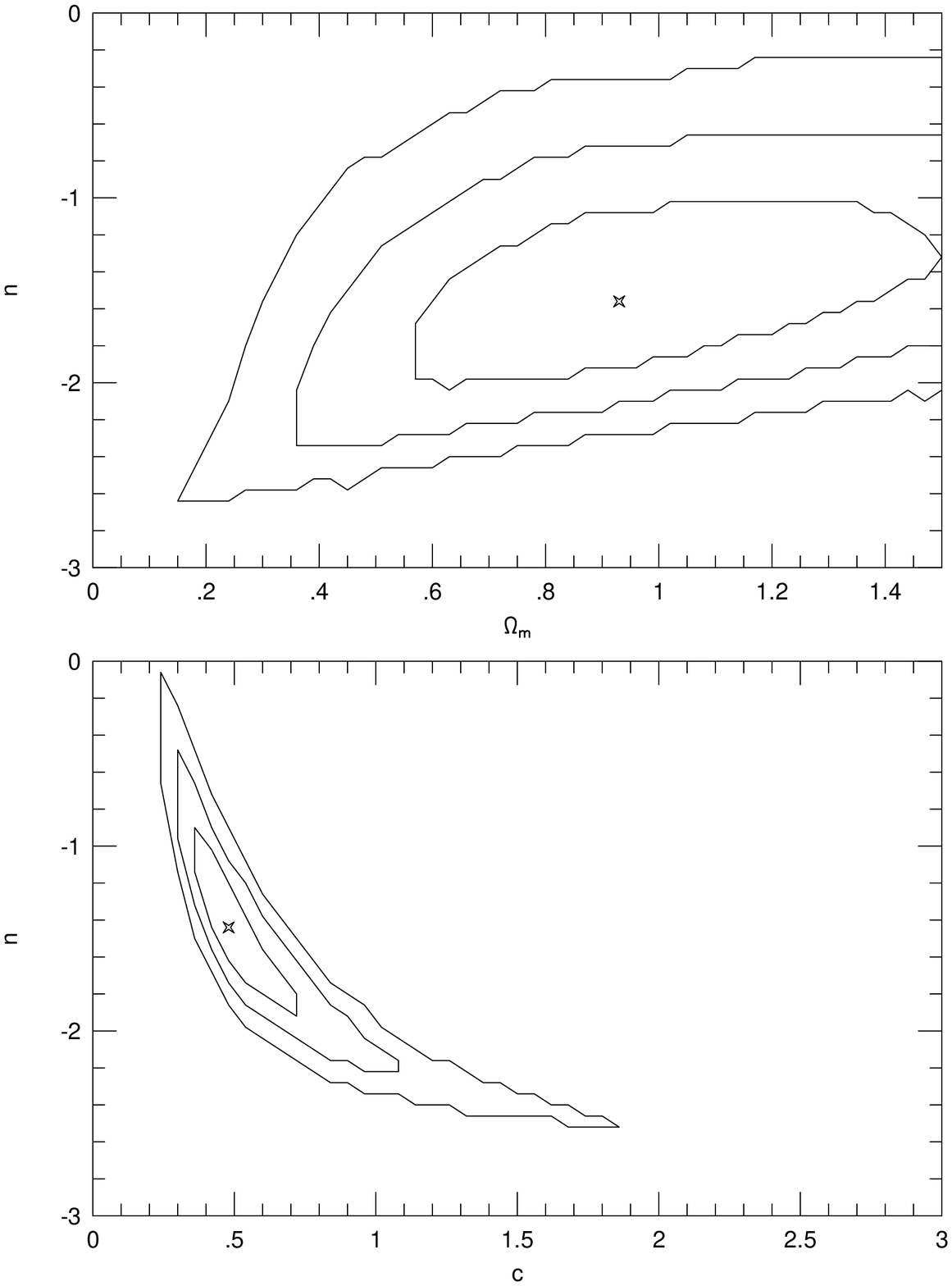]{The 1, 2, and 3 $\sigma$ credible regions of the marginalized posterior probability distributions $P(\Omega_m,n)$ (top panel) and $P(n,c)$ (bottom panel) of the fit of equation (23) to the $0.14 < z < 0.9$ revised EMSS clusters and the {\it ROSAT} BCS luminosity function (see \S3.3).\label{omfig7.ps}}

\figcaption[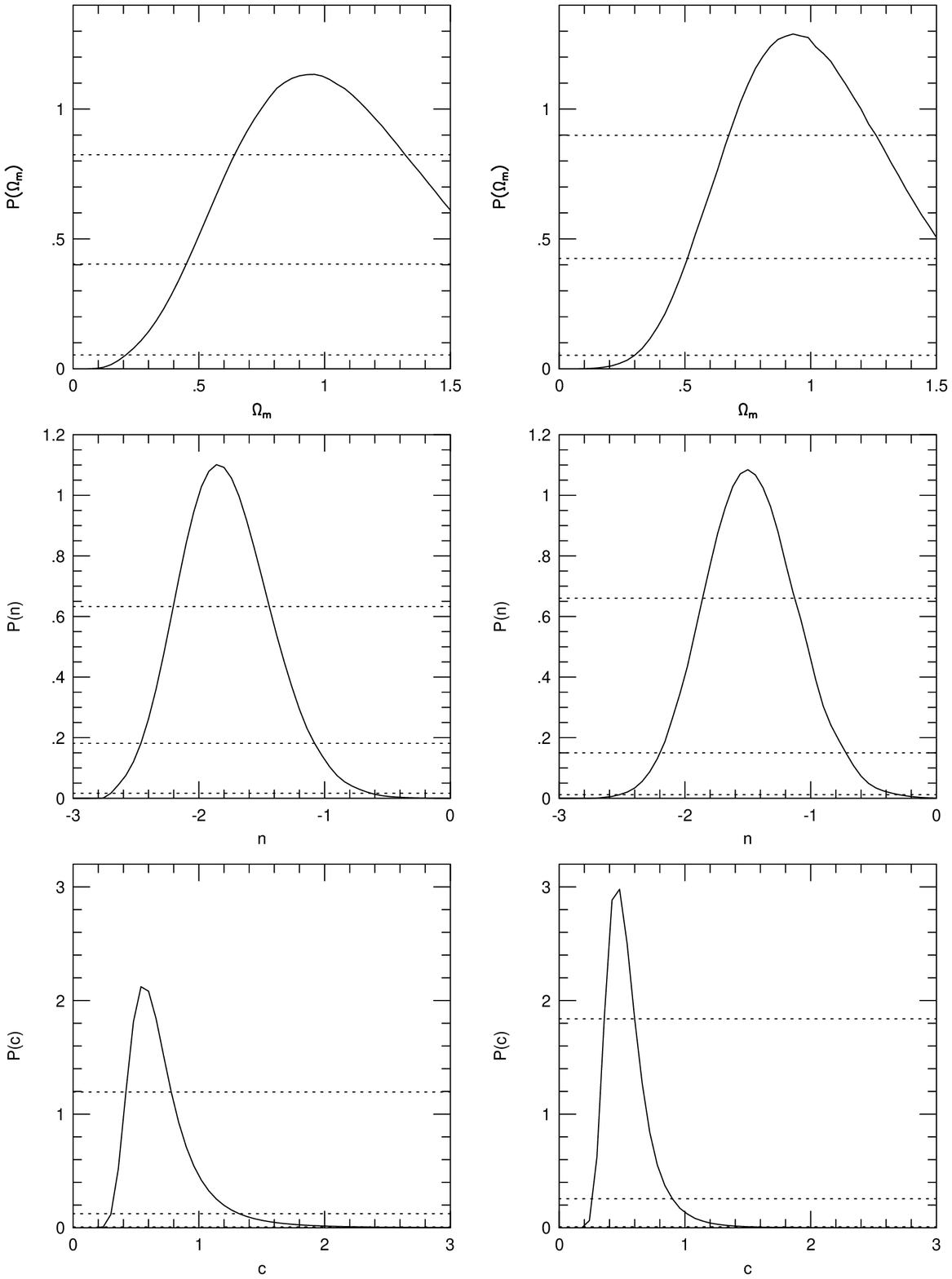]{The marginalized posterior probability distributions $P(\Omega_m)$, $P(n)$, and $P(c)$ of the fit of equation (23) to the $0.14 < z < z_u$ revised EMSS clusters and the {\it ROSAT} BCS luminosity function.  For the three left panels, $z_u = 0.6$; for the three right panels, $z_u = 0.9$.  The dotted lines mark the 1, 2, and 3 $\sigma$ credible intervals (see \S3.3).\label{omfig8.ps}}

\figcaption[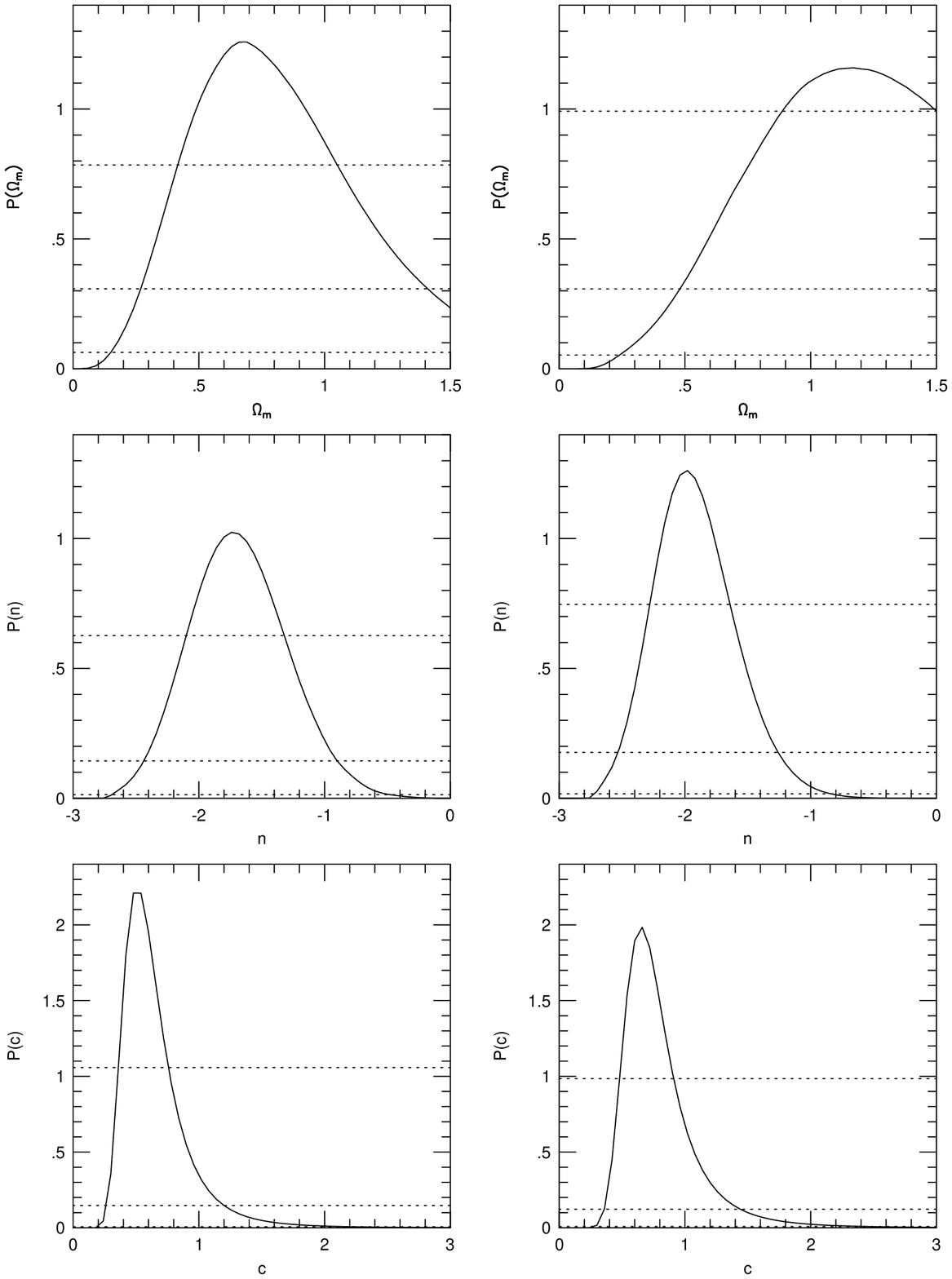]{The marginalized posterior probability distributions $P(\Omega_m)$, $P(n)$, and $P(c)$ of the fit of equation (23) to the $0.14 < z < 0.6$ revised EMSS clusters and the {\it ROSAT} BCS luminosity function.  For the three left panels, we use the $-1$ $\sigma$ value of $s$ from Reichart, Castander, \& Nichol (1998); for the three right panels, we use the $+1$ $\sigma$ value of $s$ from Reichart, Castander, \& Nichol (1998).  The dotted lines mark the 1, 2, and 3 $\sigma$ credible intervals (see \S3.3).\label{omfig9.ps}}

\clearpage

\begin{figure}[tb]
\plotone{omfig1.ps}
\end{figure}

\begin{figure}[tb]
\plotone{omfig2.ps}
\end{figure}

\begin{figure}[tb]
\plotone{omfig3.ps}
\end{figure}

\begin{figure}[tb]
\plotone{omfig4.ps}
\end{figure}

\begin{figure}[tb]
\plotone{omfig5.ps}
\end{figure}

\begin{figure}[tb]
\plotone{omfig6.ps}
\end{figure}

\begin{figure}[tb]
\plotone{omfig7.ps}
\end{figure}

\begin{figure}[tb]
\plotone{omfig8.ps}
\end{figure}

\begin{figure}[tb]
\plotone{omfig9.ps}
\end{figure}

\end{document}